\documentclass[conference, 10pt]{IEEEtran}
\IEEEoverridecommandlockouts
\usepackage{latexsym}
\usepackage{graphicx}
\usepackage{amsfonts,amssymb,amsmath}
\usepackage{hyperref}

\def\nb0{{\mathbf{0}}}
\def\nb1{{\mathbf{1}}}

\def\N{\sigma^2}

\newcommand{\ov}{\overline}
\newcommand{\bsym}{\boldsymbol}

\usepackage[acronyms,nonumberlist,nopostdot,nomain,nogroupskip]{glossaries}
\newacronym{quic}{QUIC}{Quick UDP Internet Connections}
\newacronym{3gpp}{3GPP}{3rd Generation Partnership Project}
\newacronym{adc}{ADC}{Analog to Digital Converter}
\newacronym{5g}{5G}{5th generation}
\newacronym{aimd}{AIMD}{Additive Increase Multiplicative Decrease}
\newacronym{am}{AM}{Acknowledged Mode}
\newacronym{amc}{AMC}{Adaptive Modulation and Coding}
\newacronym{aqm}{AQM}{Active Queue Management}
\newacronym{awgn}{AGWN}{Additive White Gaussian Noise}
\newacronym{afd}{AFD}{Austin Fire Department}
\newacronym{balia}{BALIA}{Balanced Link Adaptation}
\newacronym{bdp}{BDP}{Bandwidth-Delay Product}
\newacronym{bf}{BF}{Beamforming}
\newacronym{cc}{CC}{Congestion Control}
\newacronym{cdf}{CDF}{Cumulative Distribution Function}
\newacronym{cn}{CN}{Core Network}
\newacronym{cqi}{CQI}{Channel Quality Information}
\newacronym{cp}{CP}{Control Plane}
\newacronym{csirs}{CSI-RS}{Channel State Information - Reference Signal}
\newacronym{dc}{DC}{Dual Connectivity}
\newacronym{dce}{DCE}{Direct Code Execution}
\newacronym{dci}{DCI}{Downlink Control Information}
\newacronym{dl}{DL}{Downlink}
\newacronym{dmr}{DMR}{Deadline Miss Ratio}
\newacronym{dmrs}{DMRS}{DeModulation Reference Signal}
\newacronym{e2e}{E2E}{End-to-End}
\newacronym{ecn}{ECN}{Explicit Congestion Notification}
\newacronym{edf}{EDF}{Earliest Deadline First}
\newacronym{enb}{eNB}{evolved Node Base}
\newacronym{epc}{EPC}{Evolved Packet Core}
\newacronym{es}{ES}{Edge Server}
\newacronym{fdma}{FDMA}{Frequency Division Multiple Access}
\newacronym{fdd}{FDD}{Frequency Division Duplexing}
\newacronym[firstplural=Radio Access Technologies (RATs)]{rat}{RAT}{Radio Access Technology}
\newacronym{fs}{FS}{Fast Switching}
\newacronym{ftp}{FTP}{File Transfer Protocol}
\newacronym{gnb}{gNB}{Next Generation Node Base}
\newacronym{harq}{HARQ}{Hybrid Automatic Repeat reQuest}
\newacronym{hetnet}{HetNet}{Heterogeneous Network}
\newacronym{hh}{HH}{Hard Handover}
\newacronym{hol}{HOL}{Head-of-Line}
\newacronym{ia}{IA}{Initial Access}
\newacronym{imt}{IMT}{International Mobile Telecommunication}
\newacronym{iot}{IoT}{Internet of Things}
\newacronym{los}{LOS}{Line of Sight}
\newacronym{lte}{LTE}{Long Term Evolution}
\newacronym{m2m}{M2M}{Machine to Machine}
\newacronym{mac}{MAC}{Medium Access Control}
\newacronym{mc}{MC}{Multi-Connectivity}
\newacronym{mcs}{MCS}{Modulation and Coding Scheme}
\newacronym{mec}{MEC}{Mobile Edge Cloud}
\newacronym{mi}{MI}{Mutual Information}
\newacronym{mimo}{MIMO}{Multiple Input, Multiple Output}
\newacronym{mmwave}{mmWave}{millimeter wave}
\newacronym{mr}{MR}{Maximum Rate}
\newacronym{mss}{MSS}{Maximum Segment Size}
\newacronym{mtd}{MTD}{Machine-Type Device}
\newacronym{mtu}{MTU}{Maximum Transmission Unit}
\newacronym{nfv}{NFV}{Network Function Virtualization}
\newacronym{nlos}{NLOS}{Non Line of Sight}
\newacronym{nr}{NR}{New Radio}
\newacronym{ofdm}{OFDM}{Orthogonal Frequency Division Multiplexing}
\newacronym{pdcch}{PDCCH}{Physical Downlonk Control Channel}
\newacronym{pdcp}{PDCP}{Packet Data Convergence Protocol}
\newacronym{pdsch}{PDSCH}{Physical Downlink Shared Channel}
\newacronym{pdu}{PDU}{Packet Data Unit}
\newacronym{pf}{PF}{Proportional Fair}
\newacronym{pgw}{PGW}{Packet Gateway}
\newacronym{phy}{PHY}{Physical}
\newacronym{pbch}{PBCH}{Physical Broadcast Channel}
\newacronym[plural=\gls{mme}s,firstplural=Mobility Management Entities (MMEs)]{mme}{MME}{Mobility Management Entity}
\newacronym{prb}{PRB}{Physical Resource Block}
\newacronym{pss}{PSS}{Primary Synchronization Signal}
\newacronym{pucch}{PUCCH}{Physical Uplink Control Channel}
\newacronym{pusch}{PUSCH}{Physical Uplink Shared Channel}
\newacronym{rach}{RACH}{Random Access Channel}
\newacronym{ran}{RAN}{Radio Access Network}
\newacronym{red}{RED}{Robotics Emergency Deployment}
\newacronym{rf}{RF}{Radio Frequency}
\newacronym{rlc}{RLC}{Radio Link Control}
\newacronym{rlf}{RLF}{Radio Link Failure}
\newacronym{rrc}{RRC}{Radio Resource Control}
\newacronym{rrm}{RRM}{Radio Resource Management}
\newacronym{rr}{RR}{Round Robin}
\newacronym{rs}{RS}{Remote Server}
\newacronym{rsrp}{RSRP}{Reference Signal Received Power}
\newacronym{rss}{RSS}{Received Signal Strength}
\newacronym{rtt}{RTT}{Round Trip Time}
\newacronym{rw}{RW}{Receive Window}
\newacronym{rx}{RX}{Receiver}
\newacronym{sa}{SA}{standalone}
\newacronym{sack}{SACK}{Selective Acknowledgment}
\newacronym{sap}{SAP}{Service Access Point}
\newacronym{sch}{SCH}{Secondary Cell Handover}
\newacronym{scoot}{SCOOT}{Split Cycle Offset Optimization Technique}
\newacronym{sdma}{SDMA}{Spatial Division Multiple Access}
\newacronym{sinr}{SINR}{Signal to Interference plus Noise Ratio}
\newacronym{sm}{SM}{Saturation Mode}
\newacronym{snr}{SNR}{Signal to Noise Ratio}
\newacronym{son}{SON}{Self-Organizing Network}
\newacronym{ss}{SS}{Synchronization Signal}
\newacronym{srs}{SRS}{Sounding Reference Signal}
\newacronym{sss}{SSS}{Secondary Synchronization Signal}
\newacronym{tb}{TB}{Transport Block}
\newacronym{tcp}{TCP}{Transmission Control Protocol}
\newacronym{tdd}{TDD}{Time Division Duplexing}
\newacronym{tdma}{TDMA}{Time Division Multiple Access}
\newacronym{tfl}{TfL}{Transport for London}
\newacronym{tm}{TM}{Transparent Mode}
\newacronym{trp}{TRP}{Transmitter Receiver Pair}
\newacronym{tti}{TTI}{Transmission Time Interval}
\newacronym{ttt}{TTT}{Time-to-Trigger}
\newacronym{tx}{TX}{Transmitter}
\newacronym{ue}{UE}{User Equipment}
\newacronym{ul}{UL}{Uplink}
\newacronym{uml}{UML}{Unified Modeling Language}
\newacronym{um}{UM}{Unacknowledged Mode}
\newacronym{utc}{UTC}{Urban Traffic Control}
\newacronym{vm}{VM}{Virtual Machine}
\newacronym{rsrq}{RSRQ}{Reference Signal Received Quality}
\newacronym{rssi}{RSSI}{Received Signal Strength Indicator}
\newacronym{crs}{CRS}{Cell Reference Signal}
\newacronym{comp}{CoMP}{Coordinated Multi-Point}
\newacronym{cran}{C-RAN}{Cloud \acrlong{ran}}
\newacronym{ca}{CA}{Carrier Aggregation}
\newacronym{cco}{CC}{Carrier Component}
\newacronym{nsa}{NSA}{Non Stand Alone}
\newacronym{embb}{eMBB}{Enhanced Mobility Broadband}
\newacronym{bsr}{BSR}{Buffer Status Report}
\newacronym{srb}{SRB}{Service Radio Bearer}
\newacronym{scm}{SCM}{Spatial Channel Model}
\newacronym{sctp}{SCTP}{Stream Control Transmission Protocol}
\newacronym{mptcp}{MPTCP}{Multi-path TCP}
\newacronym{ietf}{IETF}{Internet Engineering Task Force}
\newacronym{os}{OS}{Operating System}
\newacronym{tls}{TLS}{Transport Layer Security}
\newacronym{rfc}{RFC}{Request for Comments}
\newacronym{http}{HTTP}{HyperText Transfer Protocol}
\newacronym{nat}{NAT}{Network Address Translation}
\newacronym{api}{API}{Application Programming Interface}
\newacronym{rto}{RTO}{Retransmission Timeout}
\newacronym{psc}{PSC}{Public Safety Communication}
\newacronym{rpgm}{RPGM}{Reference Point Group Mobility}
\newacronym{ic}{IC}{Incident Command}
\newacronym{rsu}{RSU}{Road Side Unit}
\newacronym{uav}{UAV}{unmanned aerial vehicle}
\newacronym{usv}{USV}{Unmanned Surface Vehicle}
\newacronym{uas}{UAS}{Unmanned Aerial System}
\newacronym{iab}{IAB}{Integrated Access and Backhaul}
\newacronym{qoe}{QoE}{Quality of Experience}
\newacronym{ssim}{SSIM}{Structural Similarity Index}
\newacronym{psnr}{PSNR}{Peak Signal to Noise Ratio}
\newacronym{bs}{BS}{Base Station}
\newacronym{mu}{MU}{Multiple User}
\newacronym{ag}{AG}{Air-to-Ground}
\newacronym{af}{AF}{Array Factor}
\newacronym{ula}{ULA}{Uniform Linear Array}
\newacronym{upa}{UPA}{Uniform Planar Array}
\newacronym{lcs}{LCS}{Local Coordinate System}
\newacronym{psd}{PSD}{Power Spectral Density}
\newacronym{vq}{VQ}{vector quantization}
\newacronym{a2g}{A2G}{air-to-ground}
\newacronym{em}{EM}{electromagnetic}
\newacronym{vae}{VAE}{variational autoencoder}

\def\bb0{{\mathbb{0}}}

\def\bb{{\boldsymbol{b}}}

\def\b0{{\boldsymbol{0}}}

\def\b{{\mathrm{b}}}

\def\r0{{\mathbf{0}}}

\def\bsf0{{\bm{\mathsf{0}}}}

\def\N0{{N_{\mathrm{0}}}}

\def\bsf{{\boldsymbol{s}_\mathrm{f}}}

\newcommand{\be}{\begin{equation}}
\newcommand{\ee}{\end{equation}}
\newcommand{\bal}{\begin{align}}
\newcommand{\eal}{\end{align}}

\usepackage{textcomp}
\usepackage{gensymb}
\usepackage{siunitx}
\usepackage{xcolor}
\usepackage{tikz}
\usepackage{etoolbox}
\usepackage{enumitem}
\usepackage{esvect}
\usepackage{mathtools}
\newtoggle{conf}
\toggletrue{conf}
\usetikzlibrary{chains,arrows,calc,positioning}
\usepackage{multirow}
\usepackage{comment}
\usepackage[normalem]{ulem}
\usepackage{upquote}

\usepackage{caption}
\usepackage{subcaption}

\newtoggle{conference}
\toggletrue{conference}  

\def\BibTeX{{\rm B\kern-.05em{\sc i\kern-.025em b}\kern-.08em T\kern-.1667em\lower.7ex\hbox{E}\kern-.125emX}}

\begin{document}
\title{Interpolation Techniques for Fast Channel Estimation in Ray Tracing}
\author{
 \IEEEauthorblockN{Ruibin Chen$^{*1}$, ~Jayadev Joy$^{*1}$, ~Yaqi Hu$^1$, ~Mingsheng Yin$^1$, ~Marco Mezzavilla$^2$, ~Sundeep Rangan$^1$} 
\IEEEauthorblockA{$^1$New York University, Tandon School of Engineering, Brooklyn, NY, USA\\
$^2$Dipartimento di Elettronica, Informazione e Bioingegneria (DEIB), Politecnico di Milano, Milan, Italy\\
}
\thanks{$^*$These authors contributed equally to this work.
}
}

\maketitle

\begin{abstract}
Ray tracing is increasingly utilized in wireless system simulations to estimate channel paths. In large-scale simulations with complex environments, ray tracing at high resolution can be computationally demanding. To reduce the computation, this paper presents a novel method for conducting ray tracing at a coarse set of reference points and interpolating the channels at other locations. The key insight is to interpolate the images of reflected points. In addition to the computational savings, the method directly captures the spherical nature of each wavefront enabling fast and accurate computation of channels using line-of-sight MIMO and other wide aperture techniques. Through empirical validation and comparison with exhaustive ray tracing, we demonstrate the efficacy and practicality of our approach in achieving high-fidelity channel predictions with reduced computational resources.
\end{abstract}


\IEEEpeerreviewmaketitle

\section{Introduction}

\begin{figure}[t]
    \centering
    \includegraphics[width=0.9\linewidth]{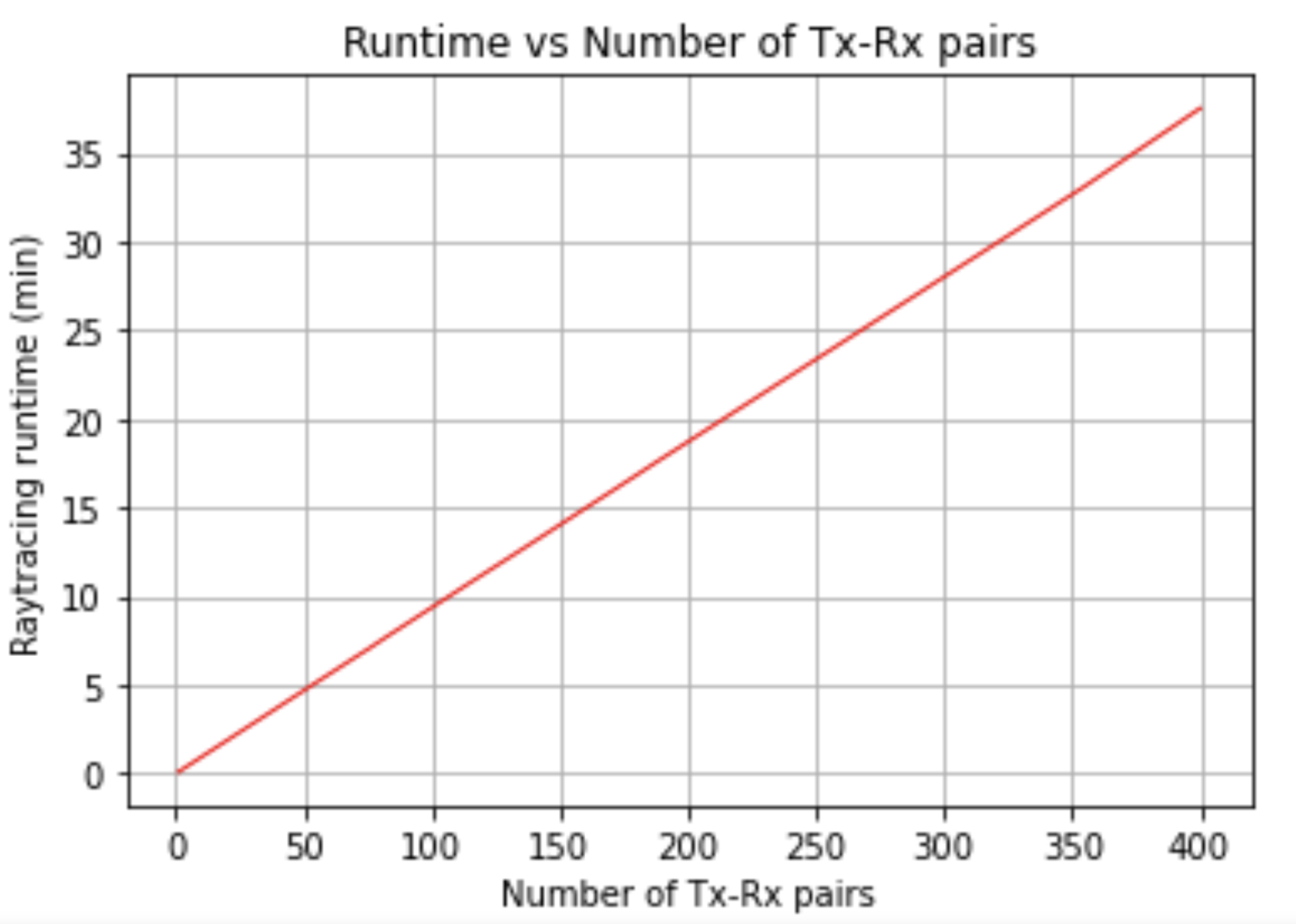}
    \caption{Ray tracing runtime complexity.}
    \label{fig:runtime}
\end{figure}

Efficient and accurate channel characterization is essential for optimizing the performance and reliability of wireless communication systems. Ray tracing has long been a fundamental technique for simulating signal propagation, providing detailed insights into the behavior of electromagnetic waves in complex environments. It is extensively utilized in applications such as site-specific planning, performance evaluation, digital twins, and other related domains. Despite its widespread adoption, conventional ray tracing methods \cite{7152831} encounter significant computational challenges, particularly when high-resolution channel characterization is required across large-scale scenarios. As illustrated in Fig. \ref{fig:runtime} (detailed of the simulation
are described below), the computational runtime increases linearly with the number of links, emphasizing the need for more efficient approaches.

To address these challenges, this paper proposes a novel approach that integrates ray tracing with interpolation techniques. Ray tracing involves tracing the paths of electromagnetic waves and capturing key channel parameters such as gains, propagation delays, and angles of arrival and departure at various receiver locations within the environment. While widely recognized as an effective tool for network planning, exhaustive ray tracing can become computationally prohibitive, especially in scenarios requiring channel estimation over a finely gridded set of reference points.

Our methodology enhances traditional ray tracing by incorporating a sophisticated reflection model derived from prior research \cite{rmmodel}. This model substitutes reflected paths with line-of-sight (LOS) paths featuring mirror image sources, addressing the limitations of plane wave models \cite{Heath} and enabling accurate representation of spherical wavefronts.

Furthermore, we employ interpolation techniques to estimate channel characteristics at target points using data from neighboring reference points. This interpolation-based framework allows for high-resolution channel predictions without the computational burden of exhaustive ray tracing at every point, thereby achieving substantial reductions in runtime while maintaining accuracy.

Through empirical validation and comparative analysis with conventional ray tracing methods, we demonstrate the effectiveness and efficiency of our approach in achieving precise channel characterization. Our research contributes to advancing channel modeling techniques, offering a practical and scalable solution for optimizing wireless communication system design and deployment. However, the proposed interpolation method may occasionally misidentify path clusters at certain extreme points, particularly in cases where the existence of LOS paths cannot be determined. Some possible solution to this challenge will be discussed in later sections.

\vspace{1mm}
\noindent
\textbf{Related Work:}
The most closely-related work is
\cite{yuan2024efficient} that proposes a method that performs ray tracing on a selected set of reference points and interpolates the remaining points using LOS parameters. This approach reduces complexity; however it primarily focuses on indoor LOS scenarios.
Interpolation is also widely used in channel estimation, for example, in OFDM systems
with time and frequency-spaced reference symbols. Linear interpolation with channel pilots is commonly employed in OFDM systems \cite{dong2007linear}, while \cite{9115838} applies an average interpolation method on a rectilinear grid of electric and magnetic field vectors. There is a large literature on 
interpolation for MIMO channel sytems as well \cite{bacci2024mmse}.  These systems, however,
interpolate in time and frequency, and do not focus
on interpolation in space as required for
ray tracing.  Nevertheless, we will see that
similar methods can be applied.

\section{Problem Formulation}

\noindent
\textbf{Discrete path model:}
For any transmit $\bsym{x}^t$ 
and receiver point $\bsym{r}^t$,
we assume that the channel can be described
by a standard discrete path model \cite{heath2018foundations} where the propagation
is given by a finite set of paths.
We let $P(\bsym{x}^t, \bsym{x}^r)$
denote the path data between 
$\bsym{x}^t$ and $\bsym{r}^t$.
In the standard plane wave model (PWA),
the path data is described by a set of parameters,
\begin{equation} \label{eq:Ppwa}
    P(\bsym{x}^t, \bsym{x}^r) = \left\{
    (g_\ell, \tau_\ell, \phi_\ell^r, \theta_\ell^r, \phi^t_\ell,
    \theta_\ell^t), \ell=1,\ldots,L 
    \right\},
\end{equation}
where $L$ is the number of paths;
and for each path $\ell$,
$g_\ell$ is its complex gain,
$(\phi_\ell^r, \theta_\ell^r)$
is the azimuth and elevation angles of arrival
(AoA, ZoA); 
$(\phi_\ell^t, \theta_\ell^t)$
is the azimuth and elevation angles of departure
(AoA, ZoA); and
$\tau_\ell$ is the path delay.

\medskip
\noindent
\textbf{Ray tracing interpolation:}
We consider a single transmitter at a location
$\ov{\bsym{x}}^t \in \mathbb{R}^3$ in 3D space, and suppose that
we have ray traced the propagation to a number
of reference receiver points, $\ov{\bsym{x}}^r_q$, $q=1,\ldots,Q$.
That is, we have obtained the path parameters
\begin{equation}
     P(\ov{\bsym{x}}^t, \ov{\bsym{x}}^r_q),
     \quad q=1,\ldots,Q.
\end{equation}
So, for each reference location, we know
the number of paths and their parameters.
Generally, the reference locations will be
placed in some coarse grid.
Our goal is, given a new receiver point $\ov{\bsym{x}}^r$, to estimate the 
path parameters $P(\ov{\bsym{x}}^t, \ov{\bsym{x}}^r)$.

\medskip
\noindent
\textbf{MIMO channel estimation:}
It is important to recognize that, once
the path parameters $P(\ov{\bsym{x}}^t,\ov{\bsym{x}}^r)$ 
are obtained, the MIMO channel matrix
can be estimated for arbitrary arrays
around $\ov{\bsym{x}}^t$ and $\ov{\bsym{x}}^r$.
The most common estimation method
is the plane wave model (PWA), which can
be described as follows:
Consider a MIMO system with $N_{tx}$
TX array elements 
at positions
$\bsym{x}^t_n$, $n=1,\ldots,N_{tx}$
in the vicinity of $\ov{\bsym{x}}^t$,
and $N_{rx}$ RX array elements
at positions
$\bsym{x}^r_m$, $m=1,\ldots,N_{rx}$
in the vicinity of $\ov{\bsym{x}}^r$.
The MIMO channel matrix is then given by
\begin{equation}
\label{eq:Hmatrix_gen}
\bsym{H}(f) = \begin{bmatrix}
H_{11}(f) & \cdots & H_{1N_{tx}}(f) \\
\vdots & \ddots & \vdots \\
H_{N_{rx}1}(f) & \cdots & H_{N_{rx}N_{tx}}(f)
\end{bmatrix}
\end{equation}
with components:
\begin{equation}
\label{eq:Hmn}
H_{mn} (f) = \sum_{\ell=1}^{L} g_\ell \exp \left[ j 2 \pi \left(\tau_\ell f_c - \frac{f d_\ell(\bsym{x}^t_m, \bsym{x}^r_n)}{v_c} \right) \right],
\end{equation}
where $f_c$ is the carrier frequency on which
the ray tracing parameters were estimated
and $d_\ell(\bsym{x}^t_m, \bsym{x}^r_n)$
is the path distance from $\bsym{x}^t_m$
to $\bsym{x}^r_n$ along path $\ell$.

In principle, this path distance 
$d_\ell(\bsym{x}^t_m, \bsym{x}^r_n)$ can be
computed by performing ray tracing
between each TX-RX element pair
$(\bsym{x}^t_m, \bsym{x}^r_n)$.
However, such an exhaustive ray tracing
would be computationally prohibitive.
For instance, an $8 \times 8$ antenna array at both the transmitter and receiver would necessitate $8^4=4096$ ray tracing experiments for a single target location. In modern 5G massive MIMO systems, which may have over 1000 array elements on one side \cite{singh2023review}, this approach becomes computationally infeasible.

Plane Wave Approximations (PWA) are widely used in industry \cite{3gpp.38.901}. The standard PWA model \cite{bohagen2009spherical} estimates the distance function $d_\ell(\bsym{x}^t_m, \bsym{x}^r_n)$ from
a linear approximation around 
the reference locations $(\ov{\bsym{x}}^t, \ov{\bsym{x}}^r)$:
\begin{equation}
\label{eq:PWA_distanceF}
d_\ell(\bsym{x}^t_m, \bsym{x}^r_n) \approx v_c \tau_\ell + (\bsym{u}_\ell^r)^\intercal(\bsym{x}^r_m - \ov{\bsym{x}}^r) +(\bsym{u}_\ell^t)^\intercal(\bsym{x}^t_m - \ov{\bsym{x}}^t),
\end{equation}
where $\tau_\ell$ is the propagation delay
on the path from the reference locations
 $(\ov{\bsym{x}}^t, \ov{\bsym{x}}^r)$ and
and $\bsym{u}_\ell^r$, $\bsym{u}_\ell^t$ are direction vectors:
\begin{equation}
\begin{split}
\label{eq:PWA_directionVector}
&\bsym{u}_\ell^r = (\cos{\phi_\ell^r}\cos{\theta_\ell^r}, \sin{\phi_\ell^r}\sin{\theta_\ell^r}, \sin{\theta_\ell^r}) \\
&\bsym{u}_\ell^t = (\cos{\phi_\ell^t}\cos{\theta_\ell^t}, \sin{\phi_\ell^t}\sin{\theta_\ell^t}, \sin{\theta_\ell^t}).
\end{split}
\end{equation}
Thus, the full MIMO matrix \eqref{eq:Hmatrix_gen}
can be estimated from the ray tracing data
at a single TX-RX pair of points
$(\ov{\bsym{x}}^t, \ov{\bsym{x}}^r)$.


\medskip
\noindent
\textbf{Reflection models for near-field
communications:}
While PWA performs well in far-field scenarios, it is less suitable for near-field communication (NFC), especially in wide-aperture MIMO systems. This limitation arises because the distance variations between Tx-Rx element pairs, negligible in far-field, become significant in NFC. The Reflection Model (RM) \cite{rmmodel} improves approximation accuracy by tracing the path routes, including reflection points, and calculating imaged Tx locations. The distance for a Tx-Rx pair is then computed as the Euclidean distance between the imaged Tx location and the Rx location.  In the RM model, 
the distance function is computed as:
\begin{equation}
\label{eq:RM_RP}
d_\ell(\bsym{x}^t_m, \bsym{x}^r_n) \approx 
\| \bsym{x}^r_m - \bsym{U}_\ell \bsym{x}^t_n \bsym{g}_\ell \|,
\end{equation}
where $\bsym{U}_\ell$ is an orthogonal matrix and $\bsym{g}_\ell$ is a shift vector.
Geometrically, the idea is that
non-line of sight distance from 
$\bsym{x}^t_m$ to $\bsym{x}^r_n$ along some path $\ell$ is identical to the LOS distance
from $\bsym{x}^r_m$ to an \emph{reflected image point} of $\bsym{x}^t_m$ given by:
\begin{equation} \label{eq:image_point}
    \bsym{z}^t_{n,\ell} = \bsym{U}_\ell \bsym{x}^t_n \bsym{g}_\ell.
\end{equation}
The image point is a translation and rotation
of the original point.  The parameters
$\bsym{U}_\ell$  and $\bsym{g}_\ell$
can be computed from the ray tracing.

It is shown in \cite{rmmodel} that
the \eqref{eq:RM_RP} model is exactly
correct in the case of infinite, specular
planar reflections and captures the spherical 
nature of each wavefront -- the critical
feature of near-field communications. 
Moreover, simulations also demonstrated
high accuracy in more complex propgation 
environments -- see \cite{rmmodel} for details.



\section{Proposed Solution}

\begin{figure}[t]
    \centering
    \includegraphics[width=1\linewidth]{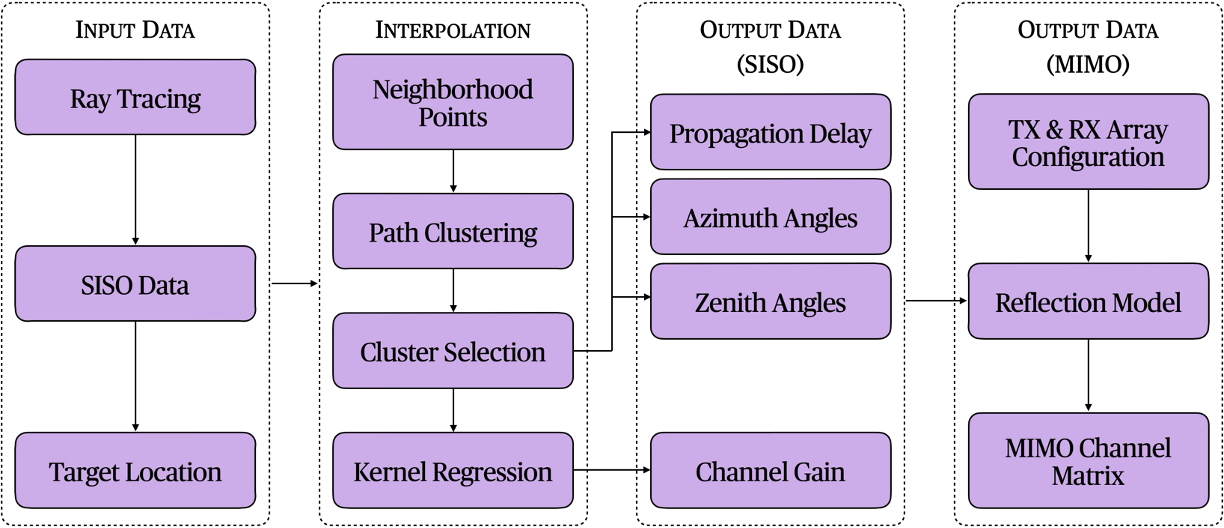}
    \caption{Overall Procedure}
    \label{fig:flowchart}
\end{figure}

We return now to our problem of estimating
path parameters  $P(\ov{\bsym{x}}^t, \ov{\bsym{x}}^r)$ from ray traced path data $P(\ov{\bsym{x}}^t, \ov{\bsym{x}}^r_q)$
at $Q$ RX reference points $q=1,\ldots,Q$.
Fig.~\ref{fig:flowchart} provides an overview of the proposed method, which begins with ray tracing data collection. The interpolation process is then applied to the data from the CSV files, followed by an evaluation of the statistical performance of the proposed method through comparisons with benchmark approaches. The details of the interpolation process are described in the following subsections.

\subsection{Data Generation}
The data at the reference points are initially acquired by performing ray tracing simulations on platforms such as Sionna within a 3D environment. This serves as the dataset for the interpolation process. Specifically, we obtain the ray traced data $P(\ov{\bsym{x}}^t, \ov{\bsym{x}}^r_q)$
at $Q$ RX reference points $q=1,\ldots,Q$.
Each ray tracing output $P(\ov{\bsym{x}}^t, \ov{\bsym{x}}^r_q)$ are the path parameters
\eqref{eq:Ppwa} and includes complex channel coefficients, path delays, elevation and azimuth angle of arrival, elevation and azimuth angle of departure.  
Additionally, path routes including detailed reflection points are also provided by most ray tracers such as open-source Sionna and the commercial Wireless Insite ray tracer \cite{remcom_wireless_insite}. These ray tracing data are stored in CSV files, allowing for efficient reuse in later stages.
Ideally, ray tracing is performed on $Q$ reference points on a relatively coarse grid
to maintain low computational cost.

\subsection{Channel Interpolation in SISO Systems}

\subsubsection{Neighborhood Reference Points}
Now consider a new target point $\ov{\bsym{x}}^r$,
not necessarily on the grid $\ov{\bsym{x}}^r_q$.
We introduce a tunable distance parameter \( d_{th} \) to identify reference points 
$\ov{\bsym{x}}^r_q$ located within the vicinity of a given target point:
\begin{equation}
    {\mathcal N} := \left\{ q ~|~ 
    \|\ov{\bsym{x}}^r - \ov{\bsym{x}}^r_q \|
    < d_{th} \| \right\},
\end{equation}
which are the indices of the reference points at most a distance of \( d_{th} \) away.

\subsubsection{Path Clustering}
We now consider all propagation paths from the source to the neighborhood reference points, replacing each reflected path with a LOS path featuring a mirror image source.
Specifically, for each reference location $q \in {\mathcal N}$ and each path $\ell$
we compute the image point:
\begin{equation} \label{eq:image_qell}
    \bsym{z}^t_{q,\ell} = \bsym{U}_{q,\ell} 
    \ov{\bsym{x}^t} + \bsym{g}_{q,\ell},
\end{equation}
where $ \bsym{U}_{q,\ell}$ and
 $\bsym{g}_{q,\ell}$ are the rotation matrix
 and translation vector for the 
 path $\ell$ to the reference image point $\bsym{x}^r_q$ as described in the RM model
 \cite{rmmodel}.
A clustering process is then performed, where paths sharing common mirror image source coordinates are grouped together. Each path cluster shares the same mirror image source coordinates. In practice, an arbitrarily small threshold is used, ensuring that paths with closely aligned image source coordinates are clustered. These clustered paths serve as the basis for the interpolation strategy.

\subsubsection{Cluster Selection}
For each path cluster \( k = 1, 2, \ldots, K \), let \( \mathcal{N}_k \) denote the set of corresponding neighborhood reference points within that cluster. We define a probability metric, \( p_k \), as follows:
\begin{equation}
\label{eq:clusterprob}
p_k = \frac{\sum_{j \in \mathcal{N}_k} \mathcal{K}(d_j)}{\sum_{i \in \mathcal{N}} \mathcal{K}(d_i)}
\end{equation}
Here, \( \mathcal{K} \) is an RBF kernel and \( d_i \) represents the distance of reference point \( i \) to the target point. This metric quantitatively captures the overall proximity of the reference points within a cluster to the target point. A value approaching \( 1 \) indicates that the cluster is in close proximity to the target point, while a value nearing \( 0 \) suggests that the cluster is situated farther away. An appropriate threshold is then set to effectively filter out the path clusters. Let the number of path clusters after filterig be $K' \leq K$

\subsubsection{Kernel Regression}
After filtering the path clusters, we apply a kernel regression based on \( d_i \) to the channel coefficients (or Fresnel-corrected coefficients) of all the reference points within a cluster. This allows us to approximate the channel coefficient at the target point from the path corresponding to the cluster under consideration.
\begin{equation}
\label{eq:clustergain}
g^{(t)}_k = \frac{\sum_{j \in \mathcal{N}_k} \mathcal{K}(d_j)g^{(j)}_{k}}{\sum_{i \in \mathcal{N}_k} \mathcal{K}(d_i)}
\end{equation}
where, for each path cluster \( k = 1, 2, \ldots, K' \), \( g^{(t)}_k \) denotes the complex channel gain of the \( k^{th} \) path at the target point, while \( g^{(j)}_k \) represents the complex channel gain of the path associated with the \( j^{th} \) reference point within the \( k^{th} \) cluster. This process is repeated for all \( K' \) clusters, yielding \( K' \) channel coefficients corresponding to the \( K' \) paths.

\subsubsection{Other Parameters}
The remaining ray tracing parameters, such as path delay, angles of arrival, and departure, are determined through straightforward geometric calculations involving \( d_i \), the mirror image source coordinates, and the target coordinates.

At the end of procedure, we have a new
set of path parameters with $K'$
paths with the delays, path gains,
and angles of arrival and departure for each path. We can also obtain the paramters for
the RM model for each path.  
These can paths can then be used 
to estimate the MIMO channel matrix,
either via a PWA model or RM model.




\section{Performance Evaluation}




\begin{figure*}[t]
     \centering
     \begin{subfigure}[b]{0.3\textwidth}
         \centering
         \includegraphics[width=\textwidth]{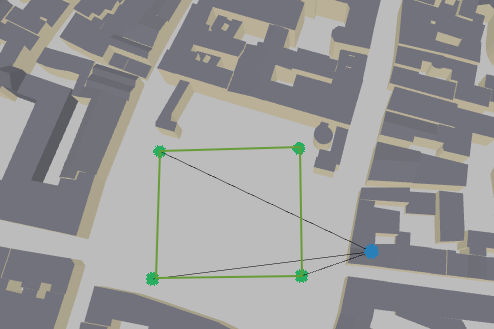}
         \caption{Most LOS}
         \label{fig:scene_LOS}
     \end{subfigure}
     \hfill
     \begin{subfigure}[b]{0.3\textwidth}
         \centering
         \includegraphics[width=\textwidth]{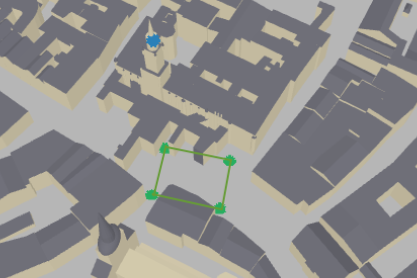}
         \caption{Partial LOS}         \label{fig:scene_partial_LOS}
     \end{subfigure}
     \hfill
     \begin{subfigure}[b]{0.3\textwidth}
         \centering
         \includegraphics[width=\textwidth]{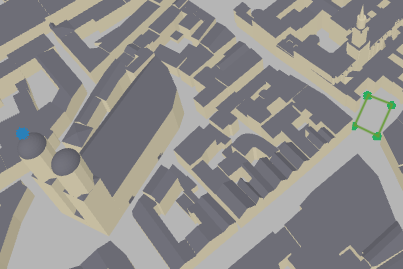}
         \caption{Total NLOS}
         \label{fig:scene_total_NLOS}
     \end{subfigure}

    \begin{subfigure}[b]{0.31\textwidth}
         \centering
         \includegraphics[width=\textwidth]{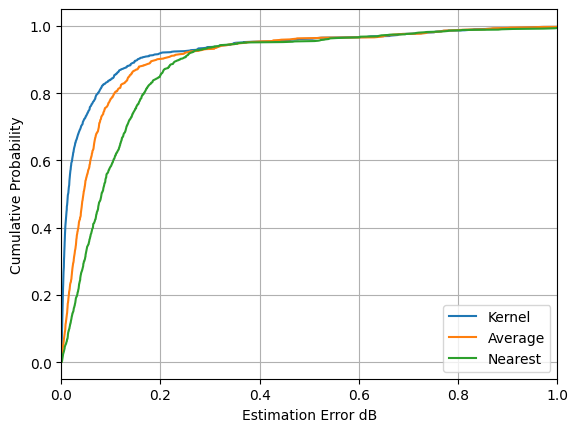}
         \caption{Power Error, Most LOS, grid = 4m}
         \label{fig:dB_LOS}
    \end{subfigure}
    \hfill
    \begin{subfigure}[b]{0.32\textwidth}
         \centering
         \includegraphics[width=\textwidth]{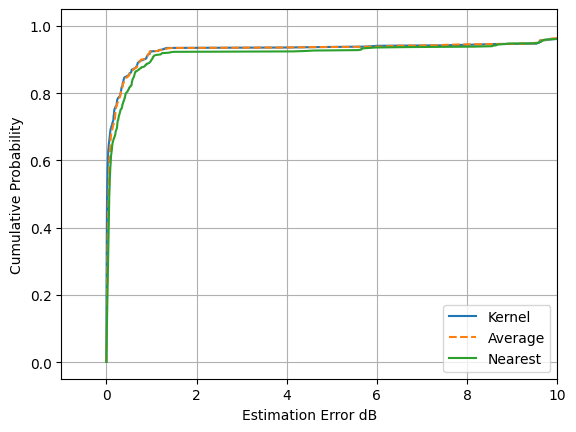}
         \caption{Power Error, Partial LOS, grid = 4m}
         \label{fig:dB_Partial_LOS}
    \end{subfigure}
    \hfill
    \begin{subfigure}[b]{0.32\textwidth}
         \centering
         \includegraphics[width=\textwidth]{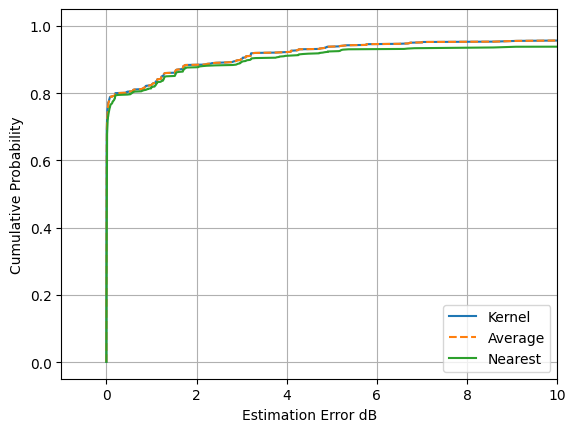}
         \caption{Power Error, NLOS, grid = 1m}
         \label{fig:dB_total_NLOS}
    \end{subfigure}

    \begin{subfigure}[b]{0.32\textwidth}
         \centering
         \includegraphics[width=\textwidth]{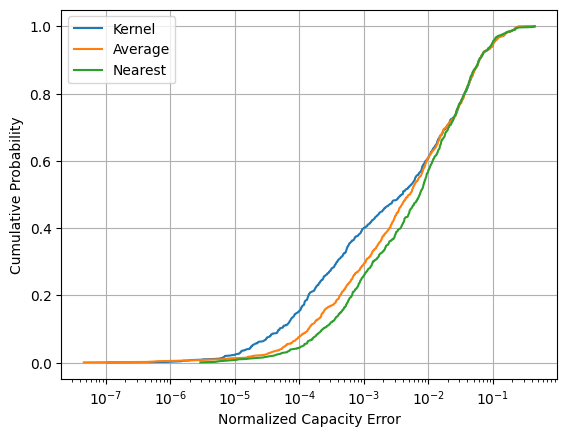}
         \caption{Capacity Error, Most LOS, grid = 4m}
         \label{fig:capacity_LOS}
    \end{subfigure}
    \hfill
    \begin{subfigure}[b]{0.32\textwidth}
         \centering
         \includegraphics[width=\textwidth]{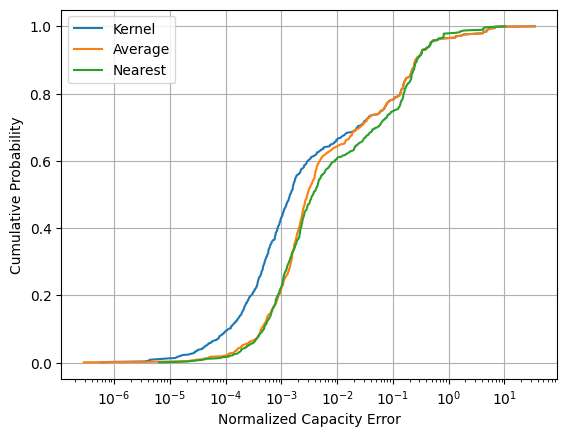}
         \caption{Capacity Error, Partial LOS, grid = 4m}
         \label{fig:capacity_Partial_LOS}
    \end{subfigure}
    \hfill
    \begin{subfigure}[b]{0.32\textwidth}
         \centering
         \includegraphics[width=\textwidth]{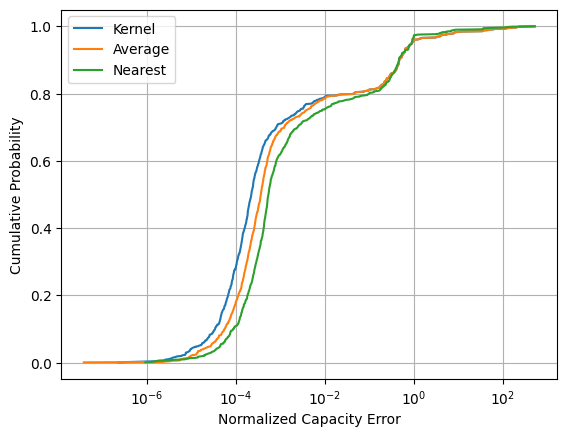}
         \caption{Capacity Error, NLOS, grid = 1m}
         \label{fig:capacity_total_NLOS}
    \end{subfigure}

        \caption{Three test scenarios with results}
        \label{fig:test_results}
\end{figure*}

\subsection{Simulation Set-Up}
Fig. \ref{fig:test_results}. illustrates a simulation setup in NVIDIA Sionna \cite{sionna}, focusing on the area surrounding Munich's Frauenkirche. The transmitter, in blue, is on the rooftop of a building. Reference receiver points, marked in green, are arranged in a rectangular grid, while target receiver points are uniformly randomized within the area. All reference and target points in this scenario are positioned at an altitude of 1.5 meters. Most reference points correspond to LOS conditions, with some exhibiting NLOS characteristics.


Fig. \ref{fig:scene_LOS}. visualizes the LOS paths for the four green corner reference points. Notably, there is no LOS path near the upper-right corner due to obstruction from nearby buildings. Fig. \ref{fig:scene_partial_LOS}. depicts a partial LOS scenario, where the upper area experiences only NLOS paths, while the lower area retains LOS connectivity. Finally, Fig. \ref{fig:scene_total_NLOS}. shows a fully NLOS scenario, where the SISO channel varies frequently and sharply.

The transmitter has three antenna arrays at the same center location, all sharing an elevation angle of approximately -10°. Each array has a distinct azimuth angle, spaced 120° apart, starting from an initial random orientation. Among the three arrays, the strongest channel between the transmitter and receiver is selected. Receiver arrays are aligned to face the transmitter's center, enhancing the LOS MIMO channel performance.

\begin{table}[ht]
\centering
\caption{Channel Estimation Simulation Parameters}
\begin{tabular}{ll}
\hline
\textbf{Parameter} & \textbf{Value} \\ \hline
\textbf{Spectrum}       & Carrier frequency: 28 GHz \\ 
\hline
\textbf{Tx antenna location} & [-35 100 30] - Mostly LOS \\
                             & [-20 -60 70] - Partial LOS \\
                             & [-210 75 105] - Fully NLOS \\

\hline
\textbf{Array Size}     & TX \& RX: 64 (8 $\times$ 8 UPA)              \\ \hline
\textbf{Antenna Spacing} &  0.14 m         \\ \hline
\textbf{Transmit Power} & 23 dBm                             \\ \hline

\textbf{Transmit Orientation} & Azimuth: 120\degree\ apart \\
    & Elevation: -10\degree \\
\hline
\textbf{Noise Figure}   & 3 dB                                         \\ \hline
\textbf{RX Array Orientation} & Facing the Tx         \\ \hline
\textbf{Sigma $\sigma$} & 1 - 3 \\ \hline
\textbf{Cluster Threshold $p_{TH}$} & 0.3 $\sim$ 0.5 \\ \hline
\end{tabular}
\label{parameters}
\end{table}

Table \ref{parameters} lists the simulation parameters used for the evaluation. The carrier frequency of 28 GHz corresponds to a wavelength of approximately 10.7$mm$, aligning with the 5G FR2 spectrum specifications defined by 3GPP.

\subsection{Total Received Power}
We evaluate the total received power gain error (in dB) to compare three interpolation strategies: simple averaging, kernel-based averaging, and nearest-point interpolation.
\begin{equation}
\label{eq:power_error}
\epsilon = \left| 10 \log_{10} \left( \sum |g_l|^2 \right) - 10 \log_{10} \left( \sum |\hat{g}_l|^2 \right) \right|
\end{equation}

In  \eqref{eq:power_error}, \( g_l \) represents the true impulse gain at the target point, obtained from ray-tracing data, while \( \hat{g}_l \) denotes the estimated impulse gain from the interpolation methods. The results shown in Fig. \ref{fig:dB_LOS}., \ref{fig:dB_Partial_LOS}., and \ref{fig:dB_total_NLOS}. depict the total received power error in dB under three test scenarios. Kernel-based interpolation performs effectively in Most LOS and Partial LOS environments but struggles in Totally NLOS cases. In the latter, smaller grid sizes are necessary for meaningful results due to the rapidly varying channel conditions, which make it challenging to predict existing paths.

\begin{figure}[t]
    \centering
    \includegraphics[width=0.7\linewidth]{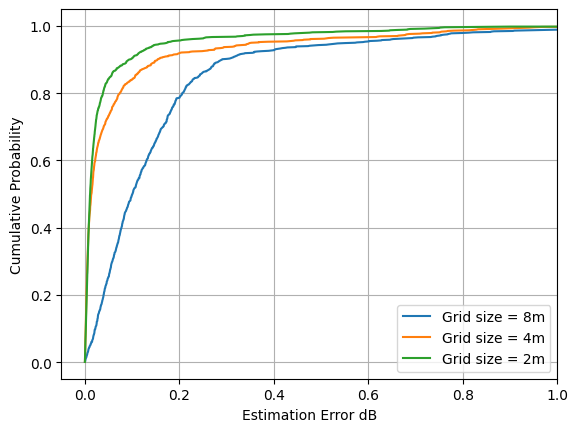}
    \caption{Power error with grid spacing in most LOS test}
    \label{fig:cdf_power_248}
\end{figure}

The grid size of the reference map impacts interpolation performance. Smaller grid sizes provide more detailed neighborhood information, leading to more accurate estimates, as illustrated in Fig. \ref{fig:cdf_power_248}. Errors for grid sizes of 2$m$ and 4$m$ are substantially lower than for 8$m$. However, denser reference maps require more ray-tracing data, with computational costs typically increasing quadratically with grid size. The choice of grid size depends on the environment, and optimizing parameters such as the kernel function's sigma value can help improve performance while allowing for larger grid sizes.

\subsection{Channel Capacity}
The channel capacity can be computed from the channel matrix \( H \) \cite{heath2018foundations}. Let the singular values of the \( H \) matrix be represented as shown in \eqref{eq:singulars}, where \( f \) is the signal frequency and \( r \) is the rank of \( H \):
\begin{equation}
\label{eq:singulars}
s(f) = (s_1(f), s_2(f), \dots, s_r(f)).
\end{equation}

The spectral efficiency \( SE(f) \) can then be estimated as:
\begin{equation}
\label{eq:se}
SE(f) = \max_{k = 1,2,\dots,r} \sum_{i=1}^k \rho \left( \frac{s_i^2 P_{tx}}{N_0 B k} \right),
\end{equation}

where \( N_0 \) is the noise power spectral density, and \( \rho (\gamma) \) is the spectral efficiency for each stream. In \eqref{eq:se}, only \( 1/k \) of the transmit power is allocated to each stream. To reflect practical scenarios, we use a 3GPP-defined model \cite{mogensen2007lte} that includes overhead, instead of the theoretical Shannon bound. In this model, \( \alpha \) (typically 0.6) is the scaling factor, and \( \rho_{\max} \) (4.8 bps/Hz) is the maximum achievable spectral efficiency:
\begin{equation}
\label{eq:rho}
\rho (\gamma) = \min\{ \alpha \log_2 (1+\gamma), \rho_{\max} \}.
\end{equation}

To evaluate interpolation techniques, we define the normalized spectral efficiency error as follows:
\begin{equation}
\label{eq:capacity_error}
\epsilon = \frac{|\hat{SE}(f) - SE(f)|}{SE(f)},
\end{equation}

where \( \hat{SE}(f) \) is the estimated spectral efficiency from interpolation, and \( SE(f) \) is the ground truth computed using ray-tracing data. The spectral efficiency under the three test cases is illustrated in Fig. \ref{fig:capacity_LOS}., \ref{fig:capacity_Partial_LOS}., and \ref{fig:capacity_total_NLOS}. Similar to the received power error, interpolation performs better in LOS MIMO environments compared to NLOS scenarios.

\begin{figure}[t]
    \centering
    \includegraphics[width=0.7\linewidth]{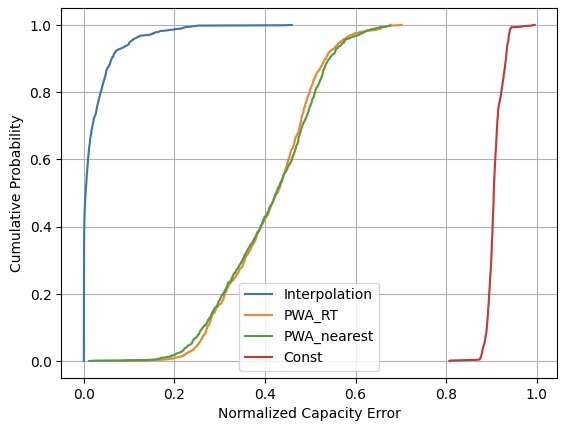}
    \caption{Capacity Error Percentage with 4m Grid}
    \label{fig:cdf_capacity_error_4m}
\end{figure}

Fig. \ref{fig:cdf_capacity_error_4m}. compares the proposed RM-based interpolation method with PWA using the nearest reference point, PWA with ray-tracing data, and the constant model. The results indicate that the proposed interpolation method significantly reduces estimation error, whereas PWA-based methods perform poorly, particularly in wide-aperture MIMO systems.

\section{Conclusion}
This work proposes a reflection model-based interpolation strategy for MIMO channel estimation using ray-tracing data. The kernel average method demonstrates superior performance compared to nearest neighbor and simple average strategies. Simulation results indicate that the proposed method is effective in most LOS and partial LOS scenarios, even with larger grid spacing (4$m$). However, its performance decreases in NLOS scenarios, where finer grid spacing (1$m$) becomes necessary. Additionally, accurately estimating points with abrupt channel variations remains challenging.

Future research on this project could explore potential solutions to address the NLOS challenges. Firstly, deep learning-based prediction techniques could be used to model the NLOS complexities, with cluster selection using deep learning expected to yield more accurate path predictions. Secondly, fine-tuning hyperparameters such as grid size, neighbor definitions, and the kernel function can optimize performance. Lastly, this study is currently constrained to simulations at 28 GHz; future research should evaluate the approach's applicability in the FR3 spectrum to ensure its generalizability.



\bibliographystyle{IEEEtran}
\bibliography{bibl}

@ARTICLE{rmmodel,
  author={Hu, Yaqi and Yin, Mingsheng and Rangan, Sundeep and Mezzavilla, Marco},
  journal={IEEE Transactions on Wireless Communications}, 
  title={Parametrization and Estimation of High-Rank Line-of-Sight MIMO Channels With Reflected Paths}, 
  year={2024},
  volume={23},
  number={4},
  pages={3808-3822},
  doi={10.1109/TWC.2023.3311735}}

@article{sionna,
    title = {Sionna: An Open-Source Library for Next-Generation Physical Layer Research},
    author = {Hoydis, Jakob and Cammerer, Sebastian and {Ait Aoudia}, Fayçal and Vem, Avinash and Binder, Nikolaus and Marcus, Guillermo and Keller, Alexander},
    year = {2022},
    month = {Mar.},
    journal = {arXiv preprint},
    online = {https://arxiv.org/abs/2203.11854}
}

@book{Heath,
place={Cambridge},
title={Foundations of MIMO Communication},
publisher={Cambridge University Press},
author={Heath Jr., Robert W. and Lozano, Angel},
year={2018}}

@ARTICLE{7152831,
  author={Yun, Zhengqing and Iskander, Magdy F.},
  journal={IEEE Access}, 
  title={Ray Tracing for Radio Propagation Modeling: Principles and Applications}, 
  year={2015},
  volume={3},
  number={},
  pages={1089-1100},
  doi={10.1109/ACCESS.2015.2453991}}

@book{heath2018foundations,
  title={Foundations of MIMO communication},
  author={Heath Jr, Robert W and Lozano, Angel},
  year={2018},
  publisher={Cambridge University Press}
}

@inproceedings{mogensen2007lte,
  title={LTE capacity compared to the Shannon bound},
  author={Mogensen, Preben and Na, Wei and Kov{\'a}cs, Istv{\'a}n Z and Frederiksen, Frank and Pokhariyal, Akhilesh and Pedersen, Klaus I and Kolding, Troels and Hugl, Klaus and Kuusela, Markku},
  booktitle={2007 IEEE 65th vehicular technology conference-VTC2007-Spring},
  pages={1234--1238},
  year={2007},
  organization={IEEE}
}

@misc{remcom_wireless_insite,
  author       = {{Remcom}},
  title        = {{Wireless InSite 3D Wireless Prediction Software}},
  howpublished = {\url{https://www.remcom.com/}},
  note         = {Accessed: Sep. 23, 2024},
  year         = {2024}
}

@article{singh2023review,
  title={A review on massive MIMO antennas for 5G communication systems on challenges and limitations},
  author={Singh, MSJ and Saleh, WSW and Abed, AT and Fauzi, MA},
  journal={Jurnal Kejuruteraan},
  volume={35},
  number={1},
  pages={95--103},
  year={2023}
}

@techreport{3gpp.38.901,
 author = {{3GPP Technique Report 38.901}},
 month = {DEC},
 title = {{Study on channel model for frequencies
from 0.5 to 100 GHz (Release 16),}},
 year = {2019}
}

@article{bohagen2009spherical,
  title={On spherical vs. plane wave modeling of line-of-sight MIMO channels},
  author={Bohagen, Frode and Orten, Pal and Oien, Geir E},
  journal={IEEE Transactions on Communications},
  volume={57},
  number={3},
  pages={841--849},
  year={2009},
  publisher={IEEE}
}

@article{dong2007linear,
  title={Linear interpolation in pilot symbol assisted channel estimation for OFDM},
  author={Dong, Xiaodai and Lu, Wu-Sheng and Soong, Anthony CK},
  journal={IEEE transactions on wireless communications},
  volume={6},
  number={5},
  pages={1910--1920},
  year={2007},
  publisher={IEEE}
}

@ARTICLE{9115838,
  author={Shikhantsov, Sergei and Thielens, Arno and Vermeeren, Günter and Demeester, Piet and Martens, Luc and Torfs, Guy and Joseph, Wout},
  journal={IEEE Journal on Selected Areas in Communications}, 
  title={Massive MIMO Propagation Modeling With User-Induced Coupling Effects Using Ray-Tracing and FDTD}, 
  year={2020},
  volume={38},
  number={9},
  pages={1955-1963},
  doi={10.1109/JSAC.2020.3000874}}

@article{bacci2024mmse,
  title={MMSE channel estimation in large-scale MIMO: Improved robustness with reduced complexity},
  author={Bacci, Giacomo and D’Amico, Antonio Alberto and Sanguinetti, Luca},
  journal={IEEE Transactions on Wireless Communications},
  year={2024},
  publisher={IEEE}
}

@inproceedings{yuan2024efficient,
  title={Efficient Ray Tracing Simulation Framework for Massive MIMO Channels},
  author={Yuan, Zhiqiang and Zhang, Jianhua and Pedersen, Gert and Fan, Wei},
  booktitle={2024 IEEE International Symposium on Antennas and Propagation and INC/USNC-URSI Radio Science Meeting (AP-S/INC-USNC-URSI)},
  pages={2187--2188},
  year={2024},
  organization={IEEE}
}

\end{document}